\documentclass[pop,jmp,preprint,graphicx,superscriptaddress]{revtex4-1}

\draft 
\usepackage{amsmath,amssymb}
\usepackage{graphicx}
\usepackage{dcolumn}
\usepackage{bm}
\usepackage{color}
\usepackage[caption=false]{subfig}
\usepackage{hyperref}   
\usepackage[all]{hypcap}
\hypersetup{
    colorlinks=true,
    linkcolor=blue,
    filecolor=magenta,      
    urlcolor=cyan,
    citecolor=red
}

\begin{document}

\preprint{AIP/123-QED}

\title{One dimensional PIC simulation of relativistic Buneman instability}
 \author{Roopendra Singh Rajawat}
\email{rupendra@ipr.res.in}
 \author{Sudip Sengupta}
\affiliation{Institute for Plasma Research, Bhat, Gandhinagar - 382428, India }
\affiliation{Homi Bhaba National Institute, Training School Complex, Anushakti Nagar, Mumbai 400085, India}
\date{\today}
\begin{abstract} 
{Spatio-temporal evolution of the relativistic Buneman instability has been investigated in one dimension using an in-house developed particle-in-cell simulation code. Starting from the excitation of the instability, its evolution has been followed numerically till its quenching and beyond. As compared to the well understood non-relativistic case, it is found that the maximum growth rate ($\gamma_{max}$) reduces due to relativistic effects and varies with $\gamma_{e0}$ and m/M as $\gamma_{max} \sim \frac{\sqrt{3}}{2\sqrt{\gamma_{e0}}}\biglb(\frac{m}{2M}\bigrb)^{1/3}$, where $\gamma_{e0}$ is Lorentz factor associated with the initial electron drift velocity ($v_{0}$) and (m/M) is the electron to ion mass ratio. Further it is observed that in contrast to the non-relativistic results[Hirose,Plasma Phys. 20, 481(1978)] at the saturation point, ratio of electrostatic field energy density ($\sum\limits_{k} |E_{k}|^{2}/8\pi$) to initial drift kinetic energy density ($W_{0}$) scales with $\gamma_{e0}$ as $\sim 1/\gamma^{2}_{e0}$. These simulation results are found to be in good agreement with that derived using fluid theory.
}
\end{abstract}
\pacs{} 
\maketitle 
\section*{INTRODUCTION}
A current carrying plasma constitutes in ideal laboratory for investigating various kinds of streaming instabilities \cite{Bret1,Bret2,Bret3,Bludman,Farley,Bunemanb}; the simplest amongst them being the electrostatic "Buneman" instability \cite{Bunemana,Buneman}, which arises when the electrons drift as a whole and the relative drift velocity between the electrons and ions exceeds the electron thermal velocity. It is associated with novel physical effects like anomalous resitivity \cite{Hirose,Yoonb,Machida}, double layer formation \cite{Nsingh,Kaw} etc. Buneman instability is of importance in many laboratory plasma experiments with intense parallel electric fields(such as in turbulent tokamaks) \cite{Tabaka,Tabak,Bandara} and in astrophysical situations with relativistic jets \cite{Nishikawa}. Recent interest in studying space time evolution and eventual saturation of Buneman instability is due to its application to a number of physical scenario's of practical interest viz. laser driven ion acceleration\cite{Yin,Albright}, strong double layer formation \cite{Nsingh,Kaw}, acceleration of charged particles \cite{Amano,Dieckmann1,Dieckmann2,Sircombe} etc.

Since the pioneering work of {Oscar Buneman} \cite{Bunemana,Buneman} a lot of work has been done to understand the linear and nonlinear evolution of Buneman instability\cite{Ichimaru, Ionson, Hirose, Ishiharaa, Ishihara, Yoona, Yoon, Jain, Hatami, Lampe, Niknam, Pavan, Che1, Che2,Hashemzadeh1}  in the non-relativistic regime \cite{briggs}. Saturation of Buneman instability in non-relativistic regime has also been studied by numerous authors \cite{Ichimaru,Ionson,Hirose}. Hirose\cite{Hirose} reported that linear saturation of Buneman instability occurs when ratio of electrostatic energy density($\sum\limits_{k} |E_{k}|^{2}/8\pi$) to initial drift kinetic energy density $W_{0}$ reaches up to $ \approx 2(m/M)^{(1/3)}$. Using quasi-linear theory, Ishihara et al \cite{Ishiharaa} derived a nonlinear dispersion relation which they verified by performing a 1-D Vlasov simulation. They further reported that linear saturation of the Buneman instability in non-relativistic regime is consistent with the Hirose's \cite{Hirose} scaling.

Recently some authors have attempted to understand the mechanism of Buneman instability in the relativistic regime. Using particle-in-cell simulation, Yin et al \cite{Yin} have found a new laser driven ion-acceleration mechanism viz. laser break-out afterburner (BOA) for production of mono-energetic ion beams in the Gev energy regime. The underlying mechanism of production of such energetic ion beams has been attributed to relativistic Buneman instability. This has been further confirmed by Albright et al \cite{Albright} by matching the results of numerical solution of dispersion relation for relativistic Buneman instability with the modes found from BOA simulation. References \cite{Dieckmann1,Dieckmann2,Sircombe} have investigated the acceleration of electrons via their interaction with electrostatic waves, driven by the relativistic Buneman instability, in a system dominated by counter-propagating proton beams. Haas\cite{Haas} et al. has investigated quantum relativistic Buneman instability using a Klein-Gordon model for the electrons and cold ions. Recently Hashemzadeh et al \cite{Hashemzadeh} have carried out 1-D particle-in-cell simulation of relativistic Buneman instability in a current carrying plasma. Their simulations show that with increase in initial electron drift velocity the growth rate of Buneman instability decreases. Although this is expected from a fluid model, a detailed comparison of the characteristics of the instability with the fluid model has not been presented. The above discussion indicates that there have been some work on relativistic Buneman instability in the recent past, but to the best of our knowledge, investigation of its evolution and saturation using particle-in-cell simulation method, and a detailed comparison of the simulation results with a fluid model have not been attempted so far.

In this paper, we study spatio-temporal evolution of relativistic Buneman instability in one dimension, using a in-house developed particle-in-cell simulation code. Starting from the excitation of the instability, its evolution is followed numerically till the saturation and beyond. We also present a comparison of our results with a weakly relativistic fluid model. For the sake of completeness, in section \ref{ge}, we present the dispersion relation for relativistic Buneman instability in the weakly relativistic limit; section \ref{gr} presents an estimate of the maximum growth rate and its comparison with the numerical solution of the dispersion relation. In section \ref{mes}, we give a brief description of the particle-in-cell simulation scheme. Section \ref{rad}, contains a presentation and discussion of our results on evolution and saturation of relativistic Buneman instability. Finally we end our paper with a summary of our results in section \ref{sum}.  
\section{Governing equation}
\subsection{LINEAR DISPERSION RELATION} \label{ge}
In this section we present a derivation of linear dispersion relation for relativistic Buneman instability. Consider a cold relativistic electron beam of density $n_{0}$ and velocity $v_{0}$ propagating through a  homogeneous background of ions of density $n_{0}$. Buneman instability occurs when relative drift velocity between electron and ion is sufficiently larger than electron thermal velocity i.e. $v_{0} \gg v_{th}$. The basic equation governing the space-time evolution of Buneman instability in 1D are as follows.\\
{The continuity equation for electrons and ions}
 {\small
\begin{equation} \label{eq:1}
 \frac{\partial n_{s}}{\partial t} + \frac{\partial \left(n_{s}{v}_{s}\right)}{\partial x} = 0    
  \end{equation}
  }
  The relativistic momentum equation for electrons and ions 
  {\small
  \begin{equation} \label{eq:2}
  \frac{\partial {p}_{s}}{\partial t} + {v}_{s}\frac{\partial \left({p}_{s}\right)}{\partial x} = \pm e{E}
  \end{equation}
	 }
	 and the Poisson equation\\
{\small
\begin{equation} \label{eq:3}
\frac{{\partial}E}{\partial x} = 4 \pi e (n_{i} - n{e})
\end{equation}
}
	 where $ s $ stands for the species(electron and ion) and $p_{s} = \frac{m_{s}v_{s}}{\sqrt{1 - (\frac{v_{s}}{c})^{2}}}$ is the relativistic momentum for species s. Here we use $m_{e} =  m$ and  $m_{i} = M$ as the rest mass of electron and ion respectively; other symbol have their usual meaning.
	 
 For electrons linearized continuity and momentum equation becomes\\
\begin{equation} \label{eq:4}
-\iota \omega \delta n_{ex} + \iota k n_{0} \delta v_{ex} + \iota k v_{0} \delta n_{ex} = 0
\end{equation}
\begin{equation} \label{eq:5}
\gamma_{e0}^{3}(-\iota \omega \delta v_{ex} + \iota k v_{0} \delta v_{ex}) = -\frac{eE}{m}
\end{equation}
where $\gamma_{e0}$ is a Lorentz factor associated with the initial electron drift velocity. Eliminating $\delta v_{ex}$ from equation (\ref{eq:4}) and (\ref{eq:5}), perturbed electron density is\\
\begin{equation} \label{eq:6}
\delta n_{ex}=\frac{-\iota e} {m \gamma_{e0}^{3}(\omega - kv_{0})^{2}}E
\end{equation} 
Again linearized continuity and momentum equation for ions can be written as\\
\begin{eqnarray} \label{eq:7}
-\iota \omega \delta n_{ix} + \iota k n_{0} \delta v_{ix} = 0
\end{eqnarray}
\begin{equation} \label{eq:8}
-\iota \omega \delta v_{ix}  = \frac{eE}{M}
\end{equation}
eliminating $\delta v_{ix}$ from equation \eqref{eq:7} and \eqref{eq:8}, gives linearized perturbed ion density as\\
\begin{equation} \label{eq:9}
\delta n_{ix} = \frac{\iota e k n_{0}}{M \omega^{2}}E
\end{equation}
Substituting from equation \eqref{eq:6} and \eqref{eq:9}, Poisson equation gives\\
\begin{equation} \label{eq:10}
\iota k E = 4 \pi (\delta n_{ix} - \delta n_{ex})
\end{equation}
Using equation (\ref{eq:6}),(\ref{eq:9}) and (\ref{eq:10}), we get the dispersion relation for Buneman instability in the weakly relativistic limit as\\
	 {\small
	 \begin{equation} \label{eq:11}
	  1 = \frac{{\omega}^{2}_{pi}}{{\omega}^{2}} + \frac{{\omega}^{2}_{pe}}{{\gamma}_{e0}^{3}({\omega} - kv_{0})^{2}} 
	 \end{equation}
	 }
\newline
where k is the wave number, $\omega_{pi} = \sqrt{\frac{4 \pi n_{0}e^{2}}{M}}$ and $\omega_{pe} = \sqrt{\frac{4 \pi n_{0}e^{2}}{m}}$ are ion and electron plasma frequency respectively.
 \subsection{ESTIMATION OF THE GROWTH RATE OF THE INSTABILITY} \label{gr}
Equation \eqref{eq:11} is a fourth order polynomial equation in $\omega$. The growth rate of the relativistic Buneman instability is given by the complex root of the  equation \eqref{eq:11} with positive imaginary part. We first give an approximate estimate of the growth rate and then compare it with that obtained using direct numerical solution of the dispersion relation. Following Haas et al\cite{Haas} we use the resonant condition $kv_{0} \approx \frac{\omega_{pe}}{\gamma^{3/2}_{e0}}$; substituting this condition in the dispersion relation and using $\omega \ll kv_{0}$ , leads to the following cubic equation in $\omega$.
\begin{equation} \label{eq:12}
{\omega}^{3}  = - \frac{m}{2M} {\gamma_{e0}}^{-3/2} {\omega}_{pe}^{3}
\end{equation}
Two complex roots of cubic equation can be written as\\
	 \begin{equation} \label{eq:14}
 	 {\omega} = \frac{(1 \pm \iota \sqrt{3})}{\sqrt{\gamma_{e0}}} \left(\frac{m}{16M} \right)^{1/3} {\omega}_{pe}
	 \end{equation}
	 The positive sign gives the growth rate of the most unstable mode as 
	 \begin{equation} \label{eq:15}
 	 {\gamma}_{max} = \frac{\sqrt{3}}{\sqrt{\gamma_{e0}}} \left(\frac{m}{16M} \right)^{1/3} {\omega}_{pe}
	 \end{equation}
Here $\gamma^{-1/2}_{e0}$ is a relativistic correction to the growth rate which explicitly shows that as $\gamma_{e0}$ increases, growth rate decreases. Most unstable k mode depends on the initial drift velocity, for example, for $kc/\omega_{pe} \approx 1$ to be the most unstable mode, initial electron drift velocity turns out to be $kv_{0}/\omega_{pe} \approx 0.65586$. Table \ref{table:1} shows the comparison between estimated (using equation \eqref{eq:15}) and numerically calculated growth rate, for the most unstable mode i.e. $kc/\omega_{pe} \approx 1$. Good matching is seen between growth rate, estimated using resonance condition (equation \eqref{eq:15}) and the growth rate obtained from numerical solution of dispersion relation.

\begin{table}
\caption{ \label{table:1}Table shows comparison between estimated and numerically calculated growth rate}
\begin{ruledtabular}
\begin{tabular}{l l l }
 M/m & $\frac{\sqrt{3}}{\sqrt{\gamma_{e0}}} \left(\frac{m}{16M} \right)^{1/3} {\omega}_{pe}$ & Numerical solution\\ 
 \hline \hline
 1836               & 0.04855 & 0.04664 \\ 
 $5\times1836$      & 0.0247  & 0.0278  \\ 
 $10\times1836$     & 0.02258 & 0.02214 \\
 $20\times1836$     & 0.0156  &  0.1553 \\
 $40\times1836$     & 0.01422 & 0.01405 \\
 \end{tabular}
\end{ruledtabular}
\end{table}

The physics underlying the resonance condition may be illustrated as follows; When electrons and ions are perturbed longitudinally by very small(linear) perturbation($\propto\exp^{\iota(kx-{\omega}t)}$), both species start to oscillates around their mean position with the frequency $\tilde{{\omega}}_{pe}$ and $\omega_{pi}$ in their respective frame of reference, where $\tilde{{\omega}}_{pe} = \frac{\omega_{pe}}{\gamma^{3/2}_{e0}}$ is the relativistically corrected electron plasma frequency and $\omega_{pi}$ is ion plasma frequency. The Doppler shifted electron oscillation can resonate with ion plasma oscillation ($\tilde{{\omega}}_{pe} - kv_{0} \approx \omega_{pi}$); in the limit of heavier ions ($\frac{\omega_{pi}}{\omega_{pe}} \rightarrow 0$), this leads to the resonance condition as $kv_{0} \approx \frac{\omega_{pe}}{\gamma^{3/2}_{e0}}$; This resonance can make ions unstable at the expense of electron drift kinetic energy and this instability is called Buneman instability. Since we get the resonance condition in the limit of heavier ions so the growth rate estimated using equation \eqref{eq:15} and the one calculated numerically come closer as the mass ratio increases. 

\section{METHOD OF SOLUTION} \label{mes}
The basic set of equations, required to study the evolution of relativistic Buneman instability in 1-D, using a particle-in-cell code\cite{Birdsall}, are the momentum and Poisson's equation. Ions are assumed to be at rest to begin with, and provide a neutralizing background while all the electrons are flowing with a single velocity $v_{e0}$. The governing equations in normalized form are 
\begin{eqnarray}
\frac{dx}{dt} = v_{s}(x,t)\\
\frac{d\gamma_{s} v_{s}}{dt} = \pm E(x,t)\\
\frac{\partial E}{\partial x} = (n_{i} - n_{e})
\end{eqnarray}
All physical quantities are used in normalized units. The normalization used are $k \rightarrow k_{L}x$, $t \rightarrow t{\omega}_{pe}$, $v \rightarrow k_{L}v/{\omega}_{pe}$, $n_{s} \rightarrow n_{s}/n_{0}$, $E \rightarrow \frac{ek_{L}E}{m{\omega}_{pe}^{2}}$, where $k_{L}$ is the wave number corresponding to the longest wavelength, which is the system length. Here $\gamma_{s}$ is a Lorentz factor and s denotes the species electrons/ions. System length is divided into 1024 equidistant cells; field quantities viz. electric field and particle density are calculated at the cell center(grid points) and particle quantities like velocities are calculated at particle positions. Each species has 102400 particles spread within 1024 grid cells, so each cell contain 100 particles. Periodic boundary conditions are used that allows only integer mode numbers as k = 1,2,3...512 in the  system. Time step is taken to be $\Delta t = 0.0196349 {\omega}_{pe}^{-1}$ ($\Delta t$ is chosen such that $\omega_{pe} \Delta t \ll 1$; we have chosen 320 time steps in a plasma period). A small thermal spread $v_{th}/v_{0} = 3 \times 10^{-4}$ is given to the electron beam in order to avoid nonphysical cold beam instability \cite{Birdsall}.  Plasma is cold($v_{th}/v_{0} \approx 0.0003$) with a very small thermal spread that fulfills the necessary condition $v_{drift} \gg v_{thermal}$, so system has favorable condition to excite Buneman instability.

In this simulation we have followed ion and electron trajectories in the self consistently generated electric field. Initially electrons and ions are placed in phase space. For a given ion and electron density, electric field is calculated on the grid points by solving Poisson's equation. Using this electric field, force is calculated on the grid points; this force is then interpolated on the particle positions. Then ion and electron momentum equations are solved using this force that yields new position and velocity. This new particle position is weighted on the grid points to evaluate density over the grid points using second order polynomial interpolation scheme which is further used to calculate the new force. This process is then repeated for thousands of time steps.
\section{RESULTS AND DISCUSSION} \label{rad}
\subsection{EVOLUTION OF RELATIVISTIC BUNEMAN INSTABILITY} 
\begin{figure}
\centering
\includegraphics[scale=0.8]{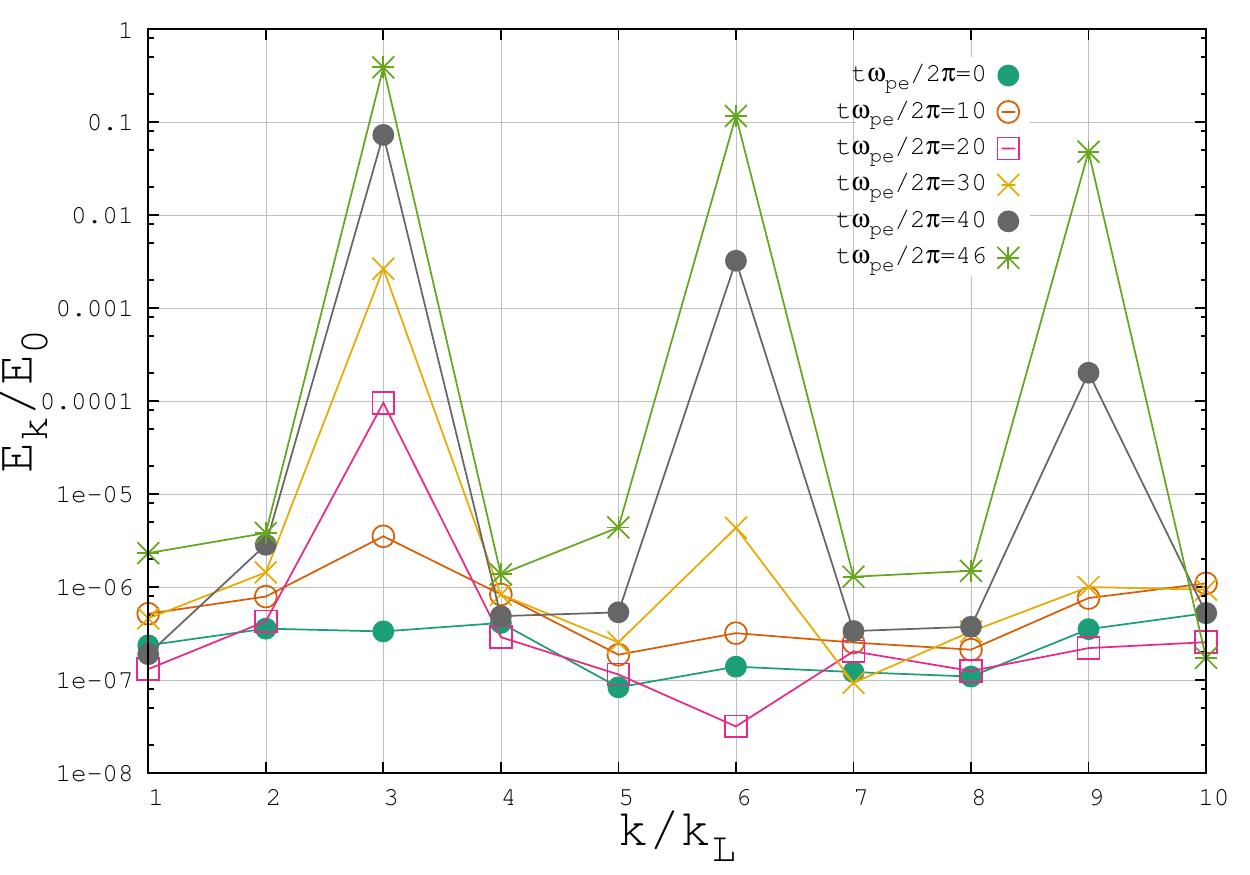}
\caption{ Evolution of k spectrum of electric field for the velocity $v_{0}/c = 0.3105$ at different time steps.}
\label{fig:fig1}
\end{figure}

\begin{figure}
\centering
\includegraphics[scale=0.8]{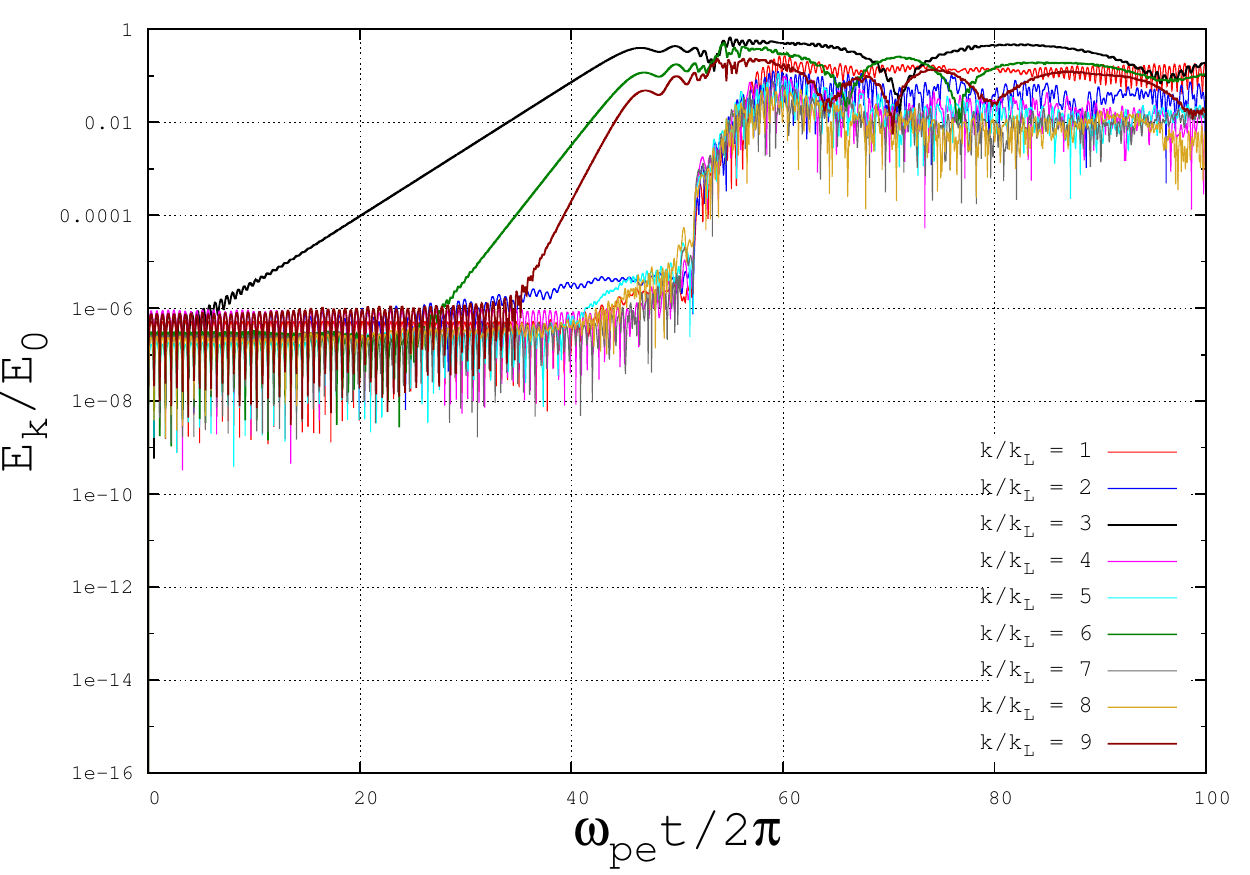}
\caption{Temporal evolution of $k^{th}$ mode of electric field for the velocity $v_{0}/c = 0.3105$.}
\label{fig:fig2}
\end{figure}

We start our simulation when the plasma is in equilibrium i.e. electrons are flowing with a single velocity, like a cold electron beam (delta function distribution) with respect to a uniform homogeneous background of ions. As time progresses, small amplitude electron, ion density and velocity oscillations evolve from background noise. Since the system is unstable and beam energy provides free energy, these small perturbations start to grow at the expense of initial beam kinetic energy density. Different modes grow at different rates. Figure (\ref{fig:fig1},\ref{fig:fig2}) show evolution of amplitude of electric field in Fourier space for the mass ratio M/m = 1836 and for initial electron drift velocity $v_{0}/c \approx 0.3105$. For these parameters, the most unstable mode number turns out to be $k/k_{L} \approx 3$. This can be seen from the resonance condition $kv_{0} \approx \frac{{\omega}_{pe}}{{\gamma}_{e0}^{3/2}} \Longrightarrow k/k_{L} \approx 3$. As expected it is observed that the most unstable growing mode supported by the system grows faster than the other modes. Temporal evolution of different Fourier modes is shown in figure (\ref{fig:fig2}). The black line shows the evolution of the most unstable mode and, green and brown lines respectively show the evolution of the first and second harmonic of the most unstable mode. Around ${\omega}_{pe}t/2{\pi} \approx 4$, the most unstable mode ($k/k_{L} = 3$) starts to evolve with growth rate $\gamma_{max} \approx 0.0529 \omega_{pe}$. It is observed that higher harmonics ($2k/k_{L} \, \& \, 3k/k_{L}$) of the most unstable mode ($k/k_{L} = 3$) appear at later times (${\omega}_{pe}t/2 \pi \approx 25$ and 35 respectively) and are found to grow at twice and thrice the growth rate of the most unstable mode. For the above parameters linear growth of relativistic Buneman instability saturates at ${\omega}_{pe}t/2 \pi \approx 46.6$.

\begin{figure}[H]
\centering
	\includegraphics[scale=0.8]{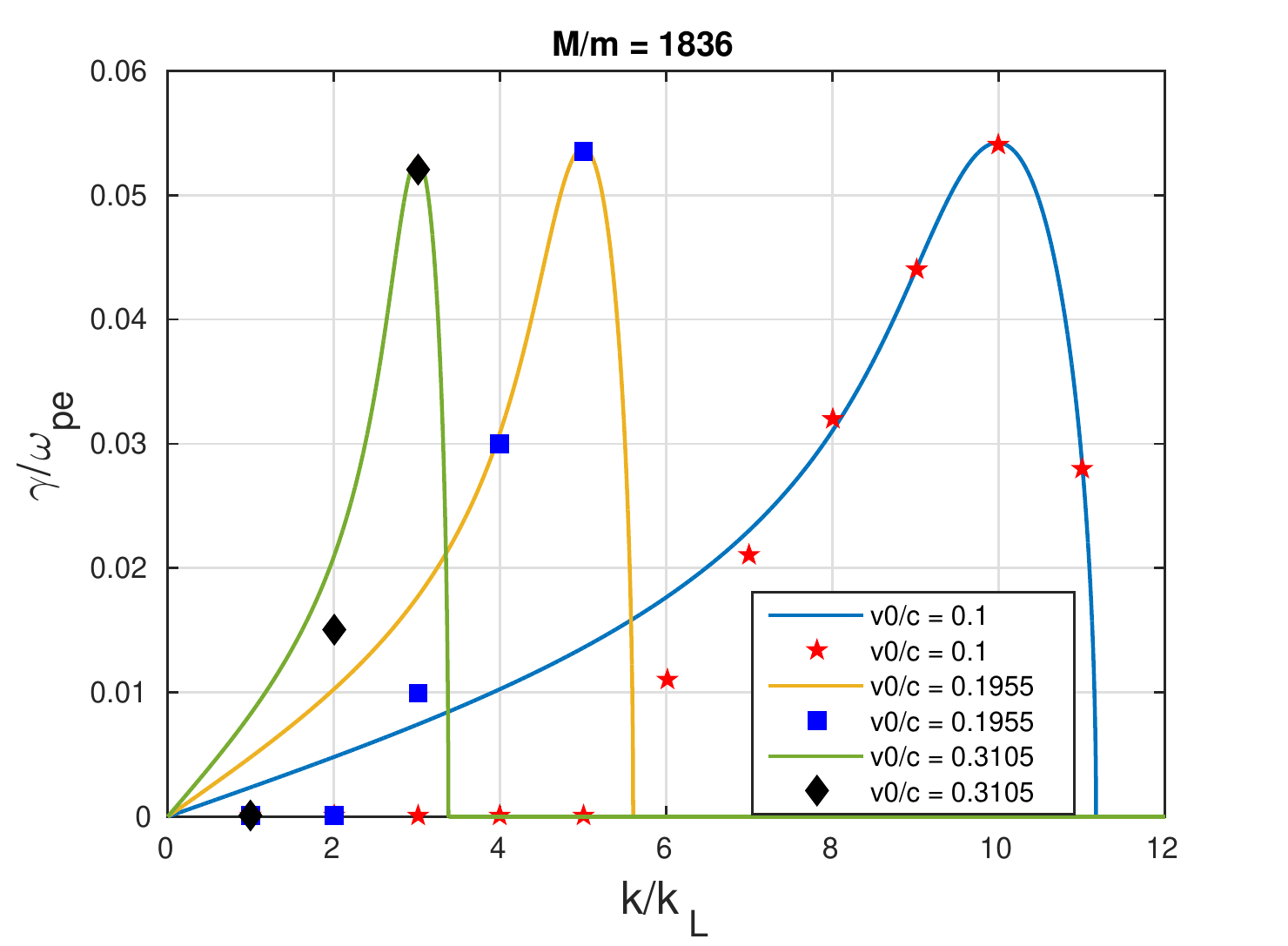}
\caption{Comparison between theory and simulation dispersion relation. Here line curves shows numerical solution of dispersion relation and dots shows growth rate taken from simulation}
\label{fig:fig3}
\end{figure}

\begin{figure}
\centering
\includegraphics[scale=0.8]{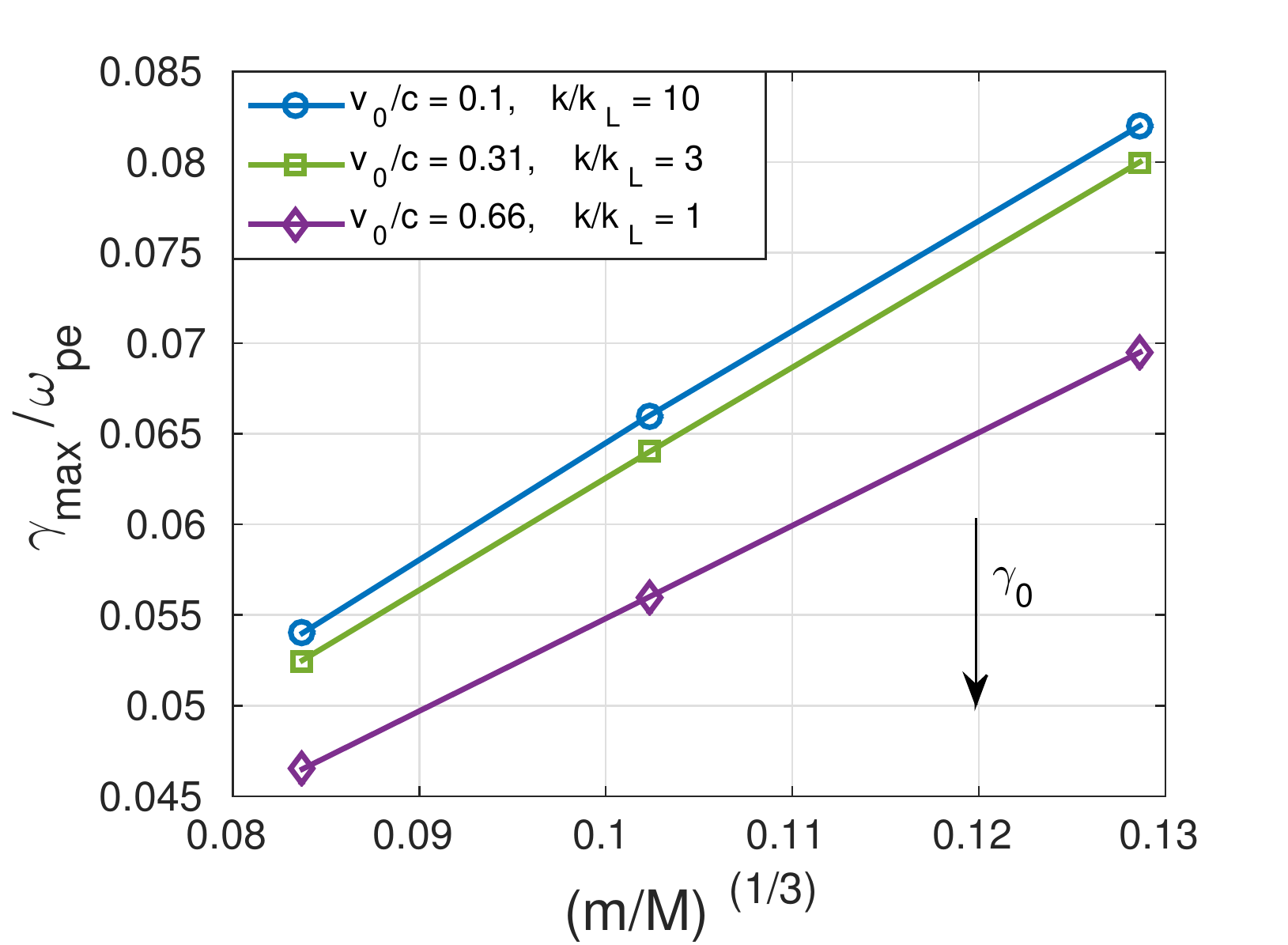}
\caption{Comparison of growth rate for different velocity with mass ratio}
\label{fig:fig5}
\end{figure}

Figure (\ref{fig:fig3}) shows the growth rate ($\gamma/\omega_{pe}$) as a function of mode number for different initial electron drift velocities and for a fixed electron to ion mass ratio. The continuous  lines are obtained by numerically solving the dispersion relation (equation \eqref{eq:11}) and the dots represent the simulation points; which shows a reasonably good match between theory and simulation. It is also clear from figure (\ref{fig:fig3}) that with the increase in velocity (relativistic effects), the peak growth rate (growth rate corresponding to the most unstable mode)
reduces for a fixed electron to ion mass ratio (m/M). This is in contrast to the non-relativistic result where the maximum growth rate corresponding to the most unstable mode number is independent of the initial electron beam drift velocity. We also note that the range of unstable mode numbers for a given initial drift velocity reduces as compared to non-relativistic case\cite{rupendra}. Figure (\ref{fig:fig5}) shows the variation of maximum growth rate with electron to ion mass ratio for different initial electron drift velocities. It is observed that the maximum growth rate ($\gamma_{max}/\omega_{pe}$) varies linearly with $(m/M)^{(1/3)}$ and decreases with increasing $v_{0}(\gamma_{e0})$ is conformity with equation \eqref{eq:15}. Thus the above results show that relativistic effects have a stabilizing influence on the Buneman instability.

\begin{figure}[H]
\centering
\subfloat[\label{fig:subfig6(1)}]{
	\includegraphics[scale=0.3]{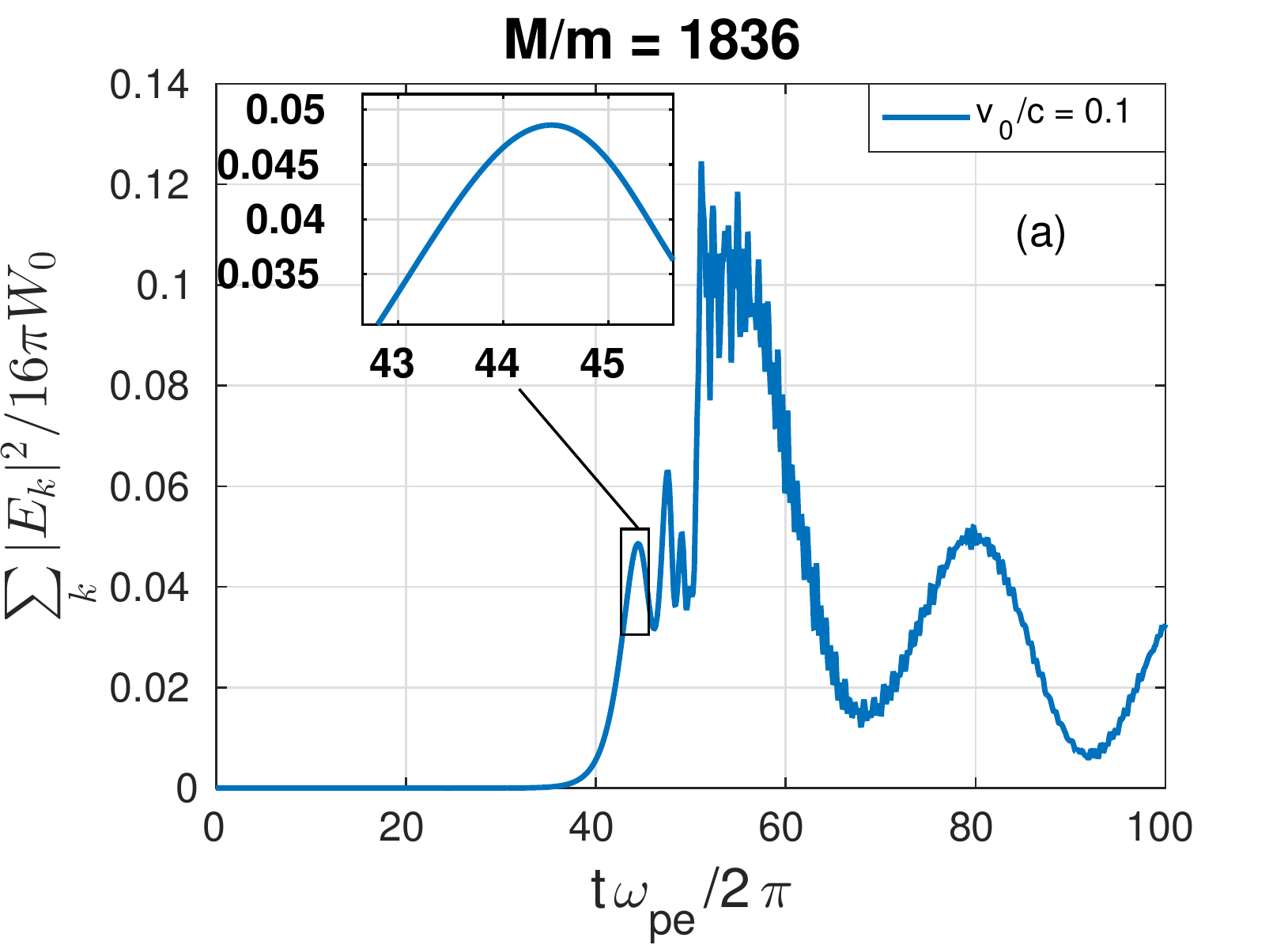}
}
\subfloat[\label{fig:subfig6(2)}]{
	\includegraphics[scale=0.3]{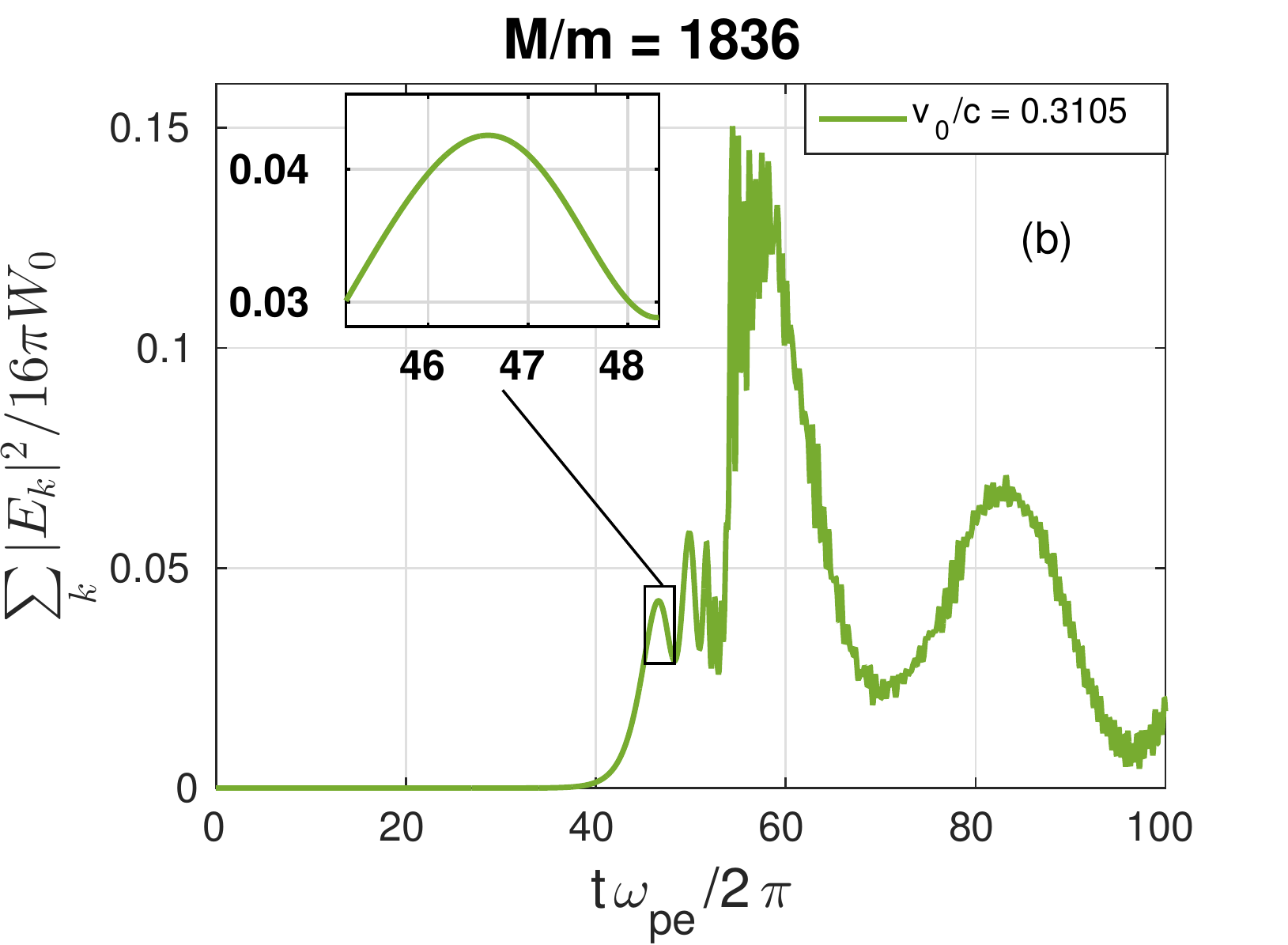}
}
\subfloat[\label{fig:subfig6(3)}]{
	\includegraphics[scale=0.3]{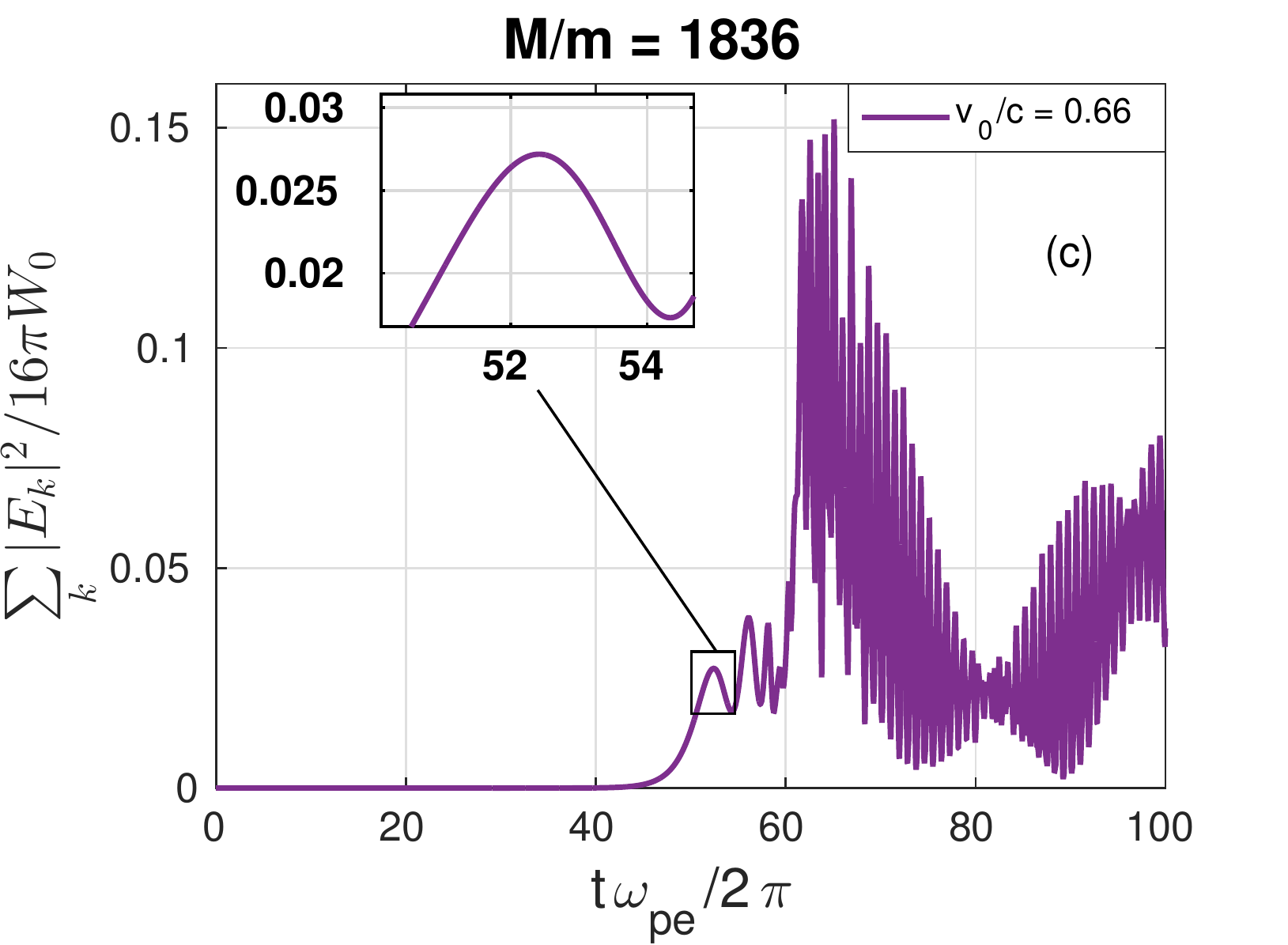}
}
\caption{Figure shows temporal evolution of $\sum\limits_{k}|E_{k}|^{2}/16 \pi W_{0}$ for different initial drift velocities (\ref{fig:subfig6(1)}) 0.1, (\ref{fig:subfig6(2)}) 0.31, (\ref{fig:subfig6(3)}) 0.66 for mass ratio M/m = 1836.}
\label{fig:fig6}
\end{figure}

\begin{figure}[H]
\centering
\subfloat[]{
	\includegraphics[scale=0.3]{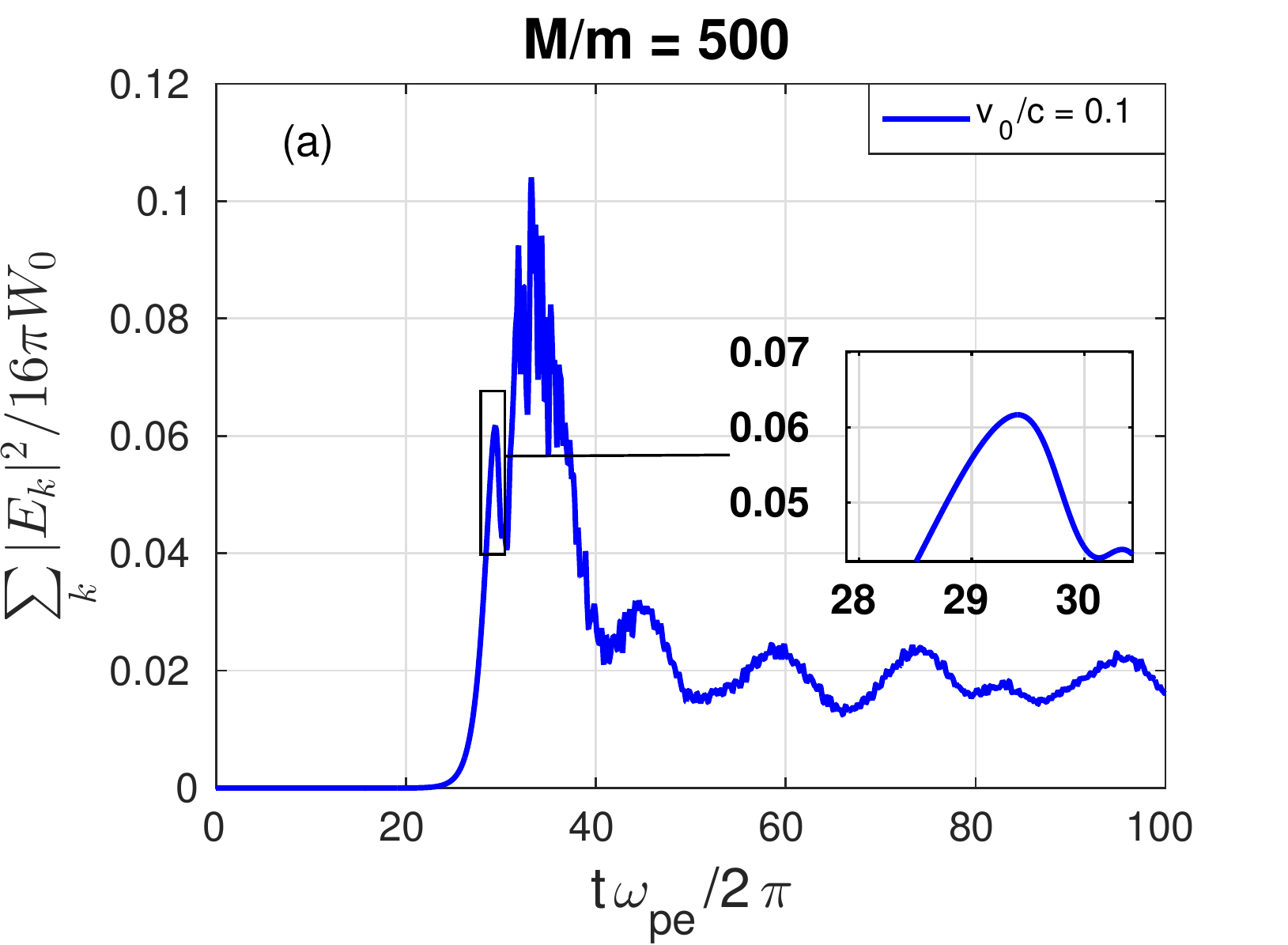}
	\label{fig:subfig7(1)}	
}
\subfloat[]{
	\includegraphics[scale=0.3]{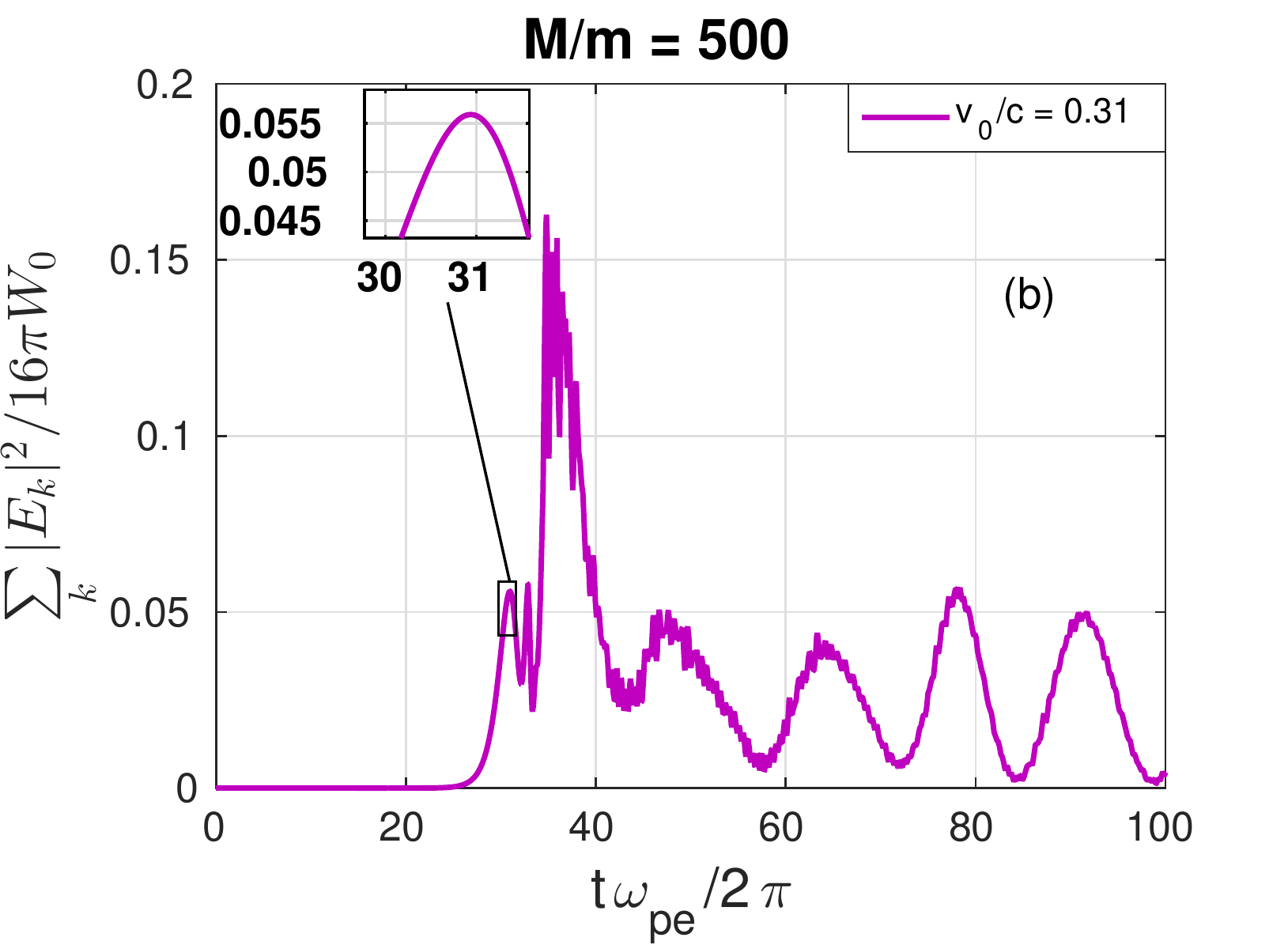}
	\label{fig:subfig7(2)}	
}
\subfloat[]{
	\includegraphics[scale=0.3]{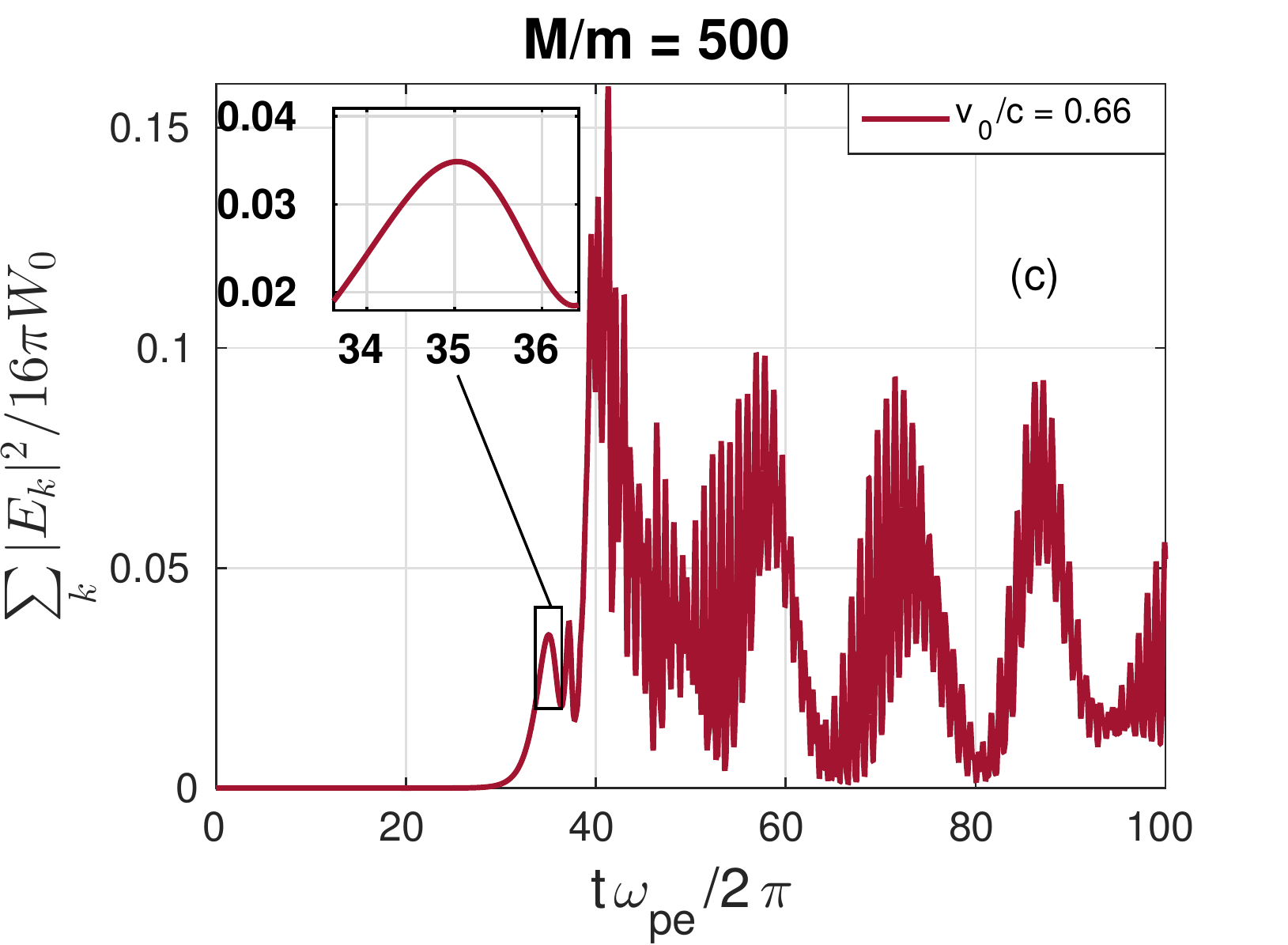}
	\label{fig:subfig7(3)}	
}
\caption{Figure shows temporal evolution of $\sum\limits_{k}|E_{k}|^{2}/16 \pi W_{0}$ for different initial drift velocities (\ref{fig:subfig7(1)}) 0.1, (\ref{fig:subfig7(2)}) 0.31, (\ref{fig:subfig7(3)}) 0.66 for mass ratio M/m = 500}
\label{fig:fig7}
\end{figure}

As mentioned in the last paragraph with the increase in initial electron drift velocity, growth rate decreases due to relativistic effects, so saturation time of instability increases. Figure (\ref{fig:fig6}) and (\ref{fig:fig7}) respectively show the temporal evolution of the electrostatic field energy for different initial electron drift velocity $v_{0}/c \approx $ 0.1, 0.3105, 0.66 for two different mass ratios M/m = 500 and 1836. These figures clearly show that as the initial electron drift velocity increases, the saturation time also increases. This is in contrast to the non-relativistic case, where the saturation time is independent of the initial electron drift velocity, and depends only on the electron to ion mass ratio (m/M). Using the saturation time for the non-relativistic \cite{rupendra} case and taking $t_{sat} \sim 1/\gamma_{max}$, we may estimate the saturation time in the relativistic case, for a fixed mass ratio (m/M) and for different initial electron drift velocities as $t^{rel}_{sat} \approx (1 + \Delta \gamma/ \gamma^{rel}_{max})t^{non-rel}_{sat}$, where $t^{rel}_{sat}$ and $t^{non-rel}_{sat}$ are the saturation times of Buneman instability for the relativistic and non-relativistic case respectively and $\Delta \gamma = \gamma^{non-rel}_{max} - \gamma^{rel}_{max}$ is the difference in growth rate of the most unstable mode in the non-relativistic and relativistic case. For example, for mass ratio M/m = 1836 and for initial electron drift velocity ($v_{0}/c = 0.3105$), the growth of the most unstable mode (in the case $k/k_{L} \approx 3$) in the non-relativistic case is $\gamma^{non-rel}_{max}/\omega_{pe} = 0.054$ (This may be estimated either by putting $\gamma_{e0} = 1$ in the relativistic dispersion relation; or by performing 1-D non-relativistic particle-in-cell simulation; our non-relativistic simulations of Buneman instability will be presented in a separate publication \cite{rupendra}) and $t^{non-rel}_{sat} \omega_{pe} /2 \pi \approx 44.46$. For the above parameters, the growth rate in the relativistic case turns out as $\gamma^{rel}_{max}/\omega_{pe} \approx 0.0525$ (estimated using equation \eqref{eq:15}). Thus the estimated saturation time in the relativistic case is $t^{rel}_{sat}\omega_{pe}/2 \pi \approx 45.73$ which is close to that observed in simulations (figure \ref{fig:subfig6(2)}). Similar estimates of $t^{rel}_{sat}$ can be made for other initial electron drift velocities and mass ratios which also show a good match with that observed in simulation.
 
\subsection{SATURATION OF THE LINEAR GROWTH OF THE INSTABILITY}
Linear saturation of the Buneman instability occurs when most unstable growing mode saturates along with its harmonics. At the saturation, electrostatic energy density shows a hiccup as shown in the figure (\ref{fig:fig6}) and (\ref{fig:fig7}) (see inset), this hiccup represents the breaking of exponential growth or linear saturation of the instability. The scaling of electrostatic field energy density at the saturation point with initial beam kinetic energy density may be derived by an analysis similar to Hirose's \cite{Hirose} for the non-relativistic case. We first reproduce Hirose's \cite{Hirose} argument here for the sake of continuity. Analysis of non-relativistic Buneman instability shows that for a given initial electron drift velocity $v_{0}$, the growth rate ($\gamma/\omega_{pe}$) maximizes at the resonant wave number given by $kv_{0} \sim \omega_{pe}$ and sharply drops for small changes in the drift velocity; the width of the $\gamma/\omega_{pe}$ vs $kv_{0}/\omega_{pe}$ curve scales with electron to ion mass ratio as $\Delta(kv_{0}/\omega_{pe}) \sim (m/M)^{1/3}$. Thus any small change in the electron drift velocity drastically reduces the growth rate resulting in quenching of the instability. This idea has been used by Hirose\cite{Hirose} to estimate the saturated electrostatic field energy density for a given initial beam kinetic energy density. Based on a quasi-linear calculation, Hirose \cite{Hirose} has shown that the ratio of $k \Delta v_{0}/\omega_{pe}$ (where "k" is the resonant wave number and $\Delta v_{0}$ is the difference between the drift velocity at the saturation time and the initial time) and $\Delta (kv_{0}/\omega_{pe})$ (the width of $\gamma/\omega_{pe}$ vs $kv_{0}/\omega_{pe}$  curve) is given by
\begin{equation} \label{eq:16}
\frac{k \Delta v_{0}}{\Delta(k v_{0})} \approx \sum\limits_{k} \frac{|E_{k}|^{2}}{16 \pi W_{0}} \left(\frac{M}{m}\right)^{1/3} \approx \frac{Field \hspace{1.5mm} energy \hspace{1.5mm} density}{Initial \hspace{1.5mm} beam \hspace{1.5mm} kinetic \hspace{1.5mm} energy \hspace{1.5mm} density} \left(\frac{M}{m}\right)^{1/3}
\end{equation}
where $W_{0}$ is the initial beam kinetic energy density. In the non-relativistic case Hirose \cite{Hirose} argued that this ratio at the saturation time should be of order unity and therefore the  electrostatic field energy density at the saturation point scales linearly with initial beam kinetic energy density, with a slope which depends on electron to ion mass ratio as $\left(m/M\right)^{1/3}$ (we have verified this by performing a 1D non-relativistic particle-in-cell simulation of Buneman instability \cite{rupendra}).

\begin{figure}
\centering
\includegraphics[scale=0.5]{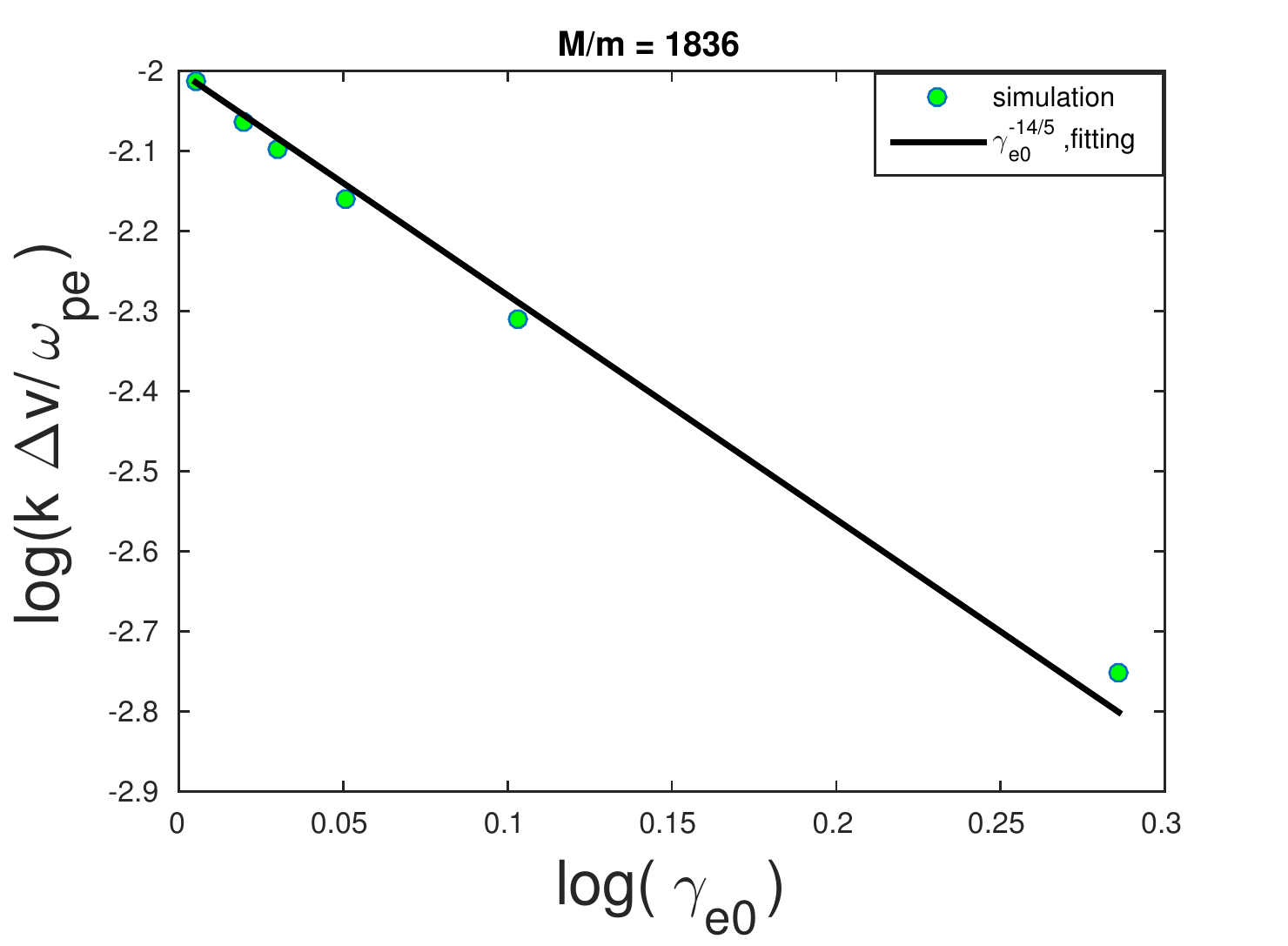}
\label{fig:fig9}
\caption{Figure shows scaling of $\frac{k\Delta v}{\omega_{pe}}$ with $\gamma_{e0}$ in log-log plot, it follows $\sim \gamma^{-14/5}_{e0}$ scaling.}
\end{figure}

\begin{figure}[H]
\centering
\subfloat[]{
	\includegraphics[scale=0.5]{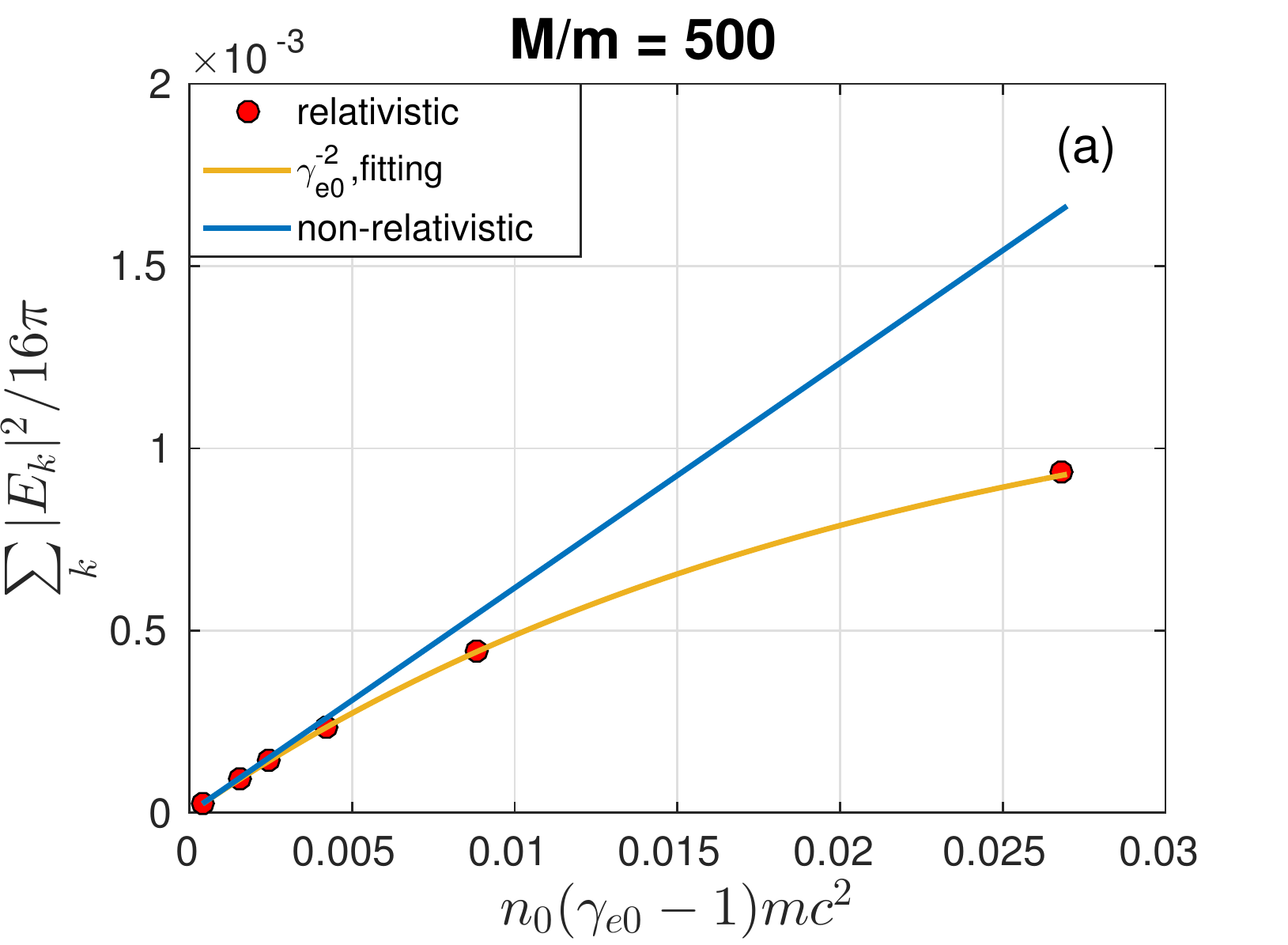}
	\label{fig:fig10(1)}	
	}
\subfloat[]{
	\includegraphics[scale=0.5]{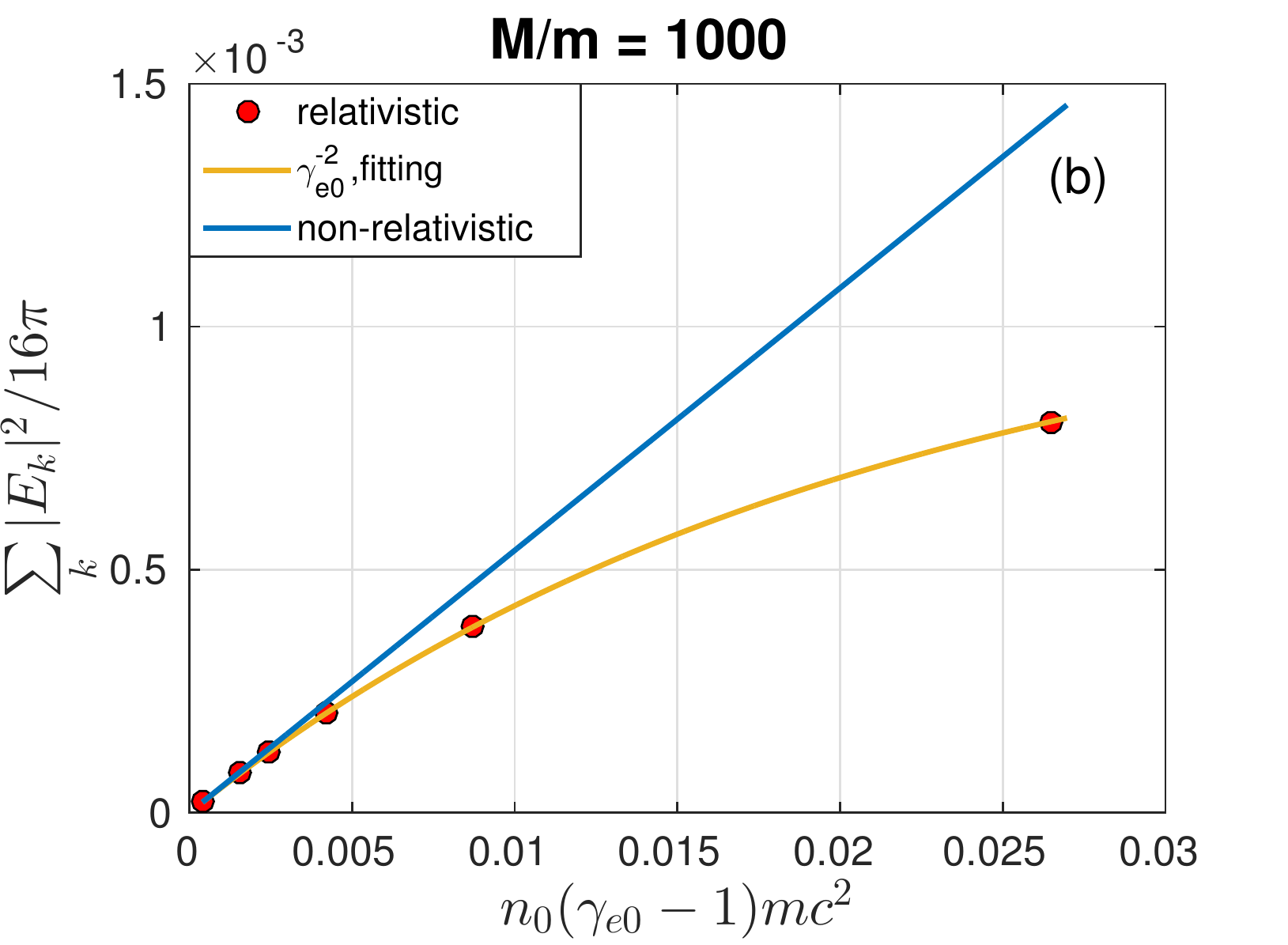}
	\label{fig:fig10(2)}
	}\\
\subfloat[]{
	\includegraphics[scale=0.5]{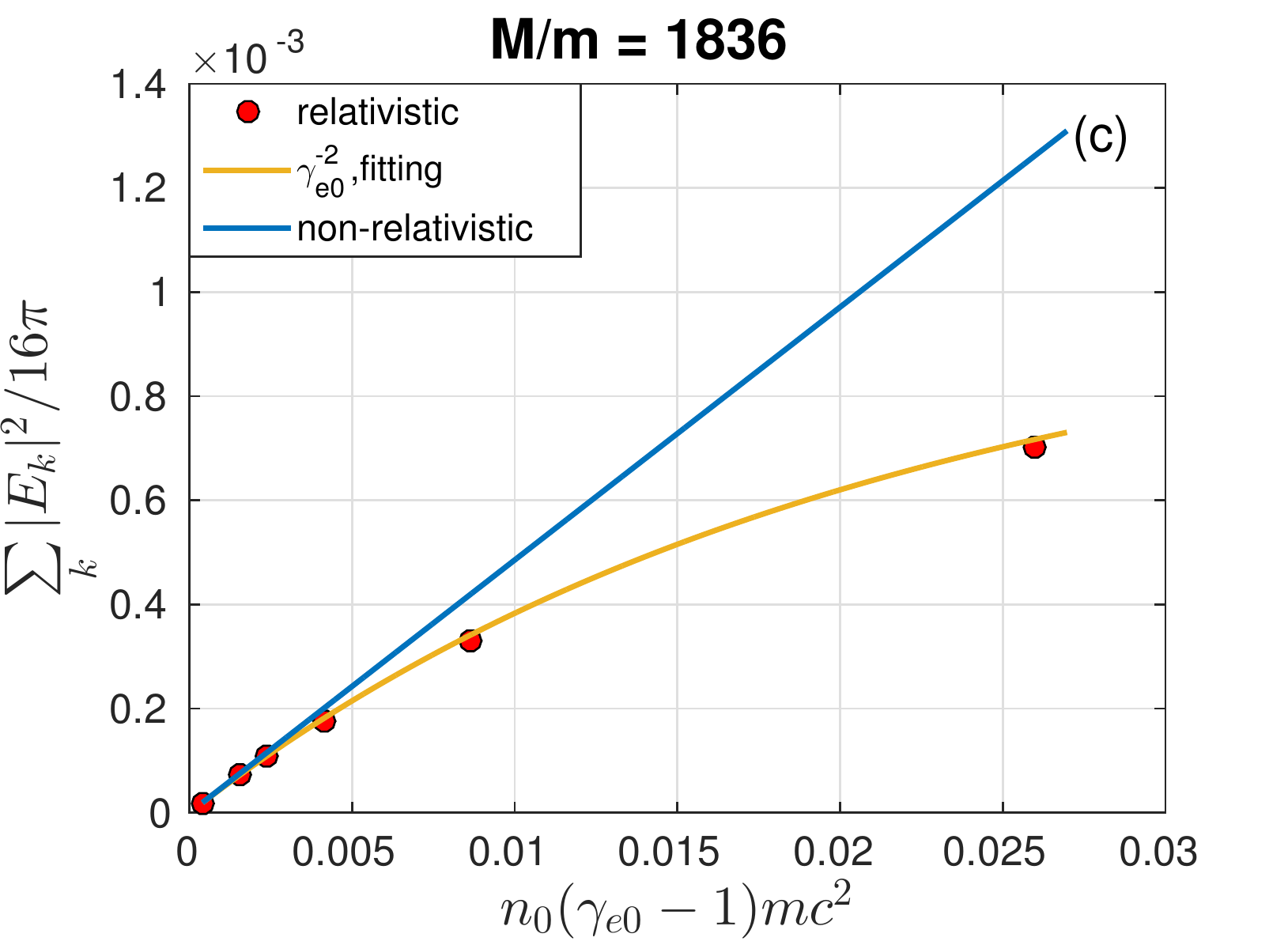}
	\label{fig:fig10(3)}
	}
\subfloat[]{
	\includegraphics[scale=0.5]{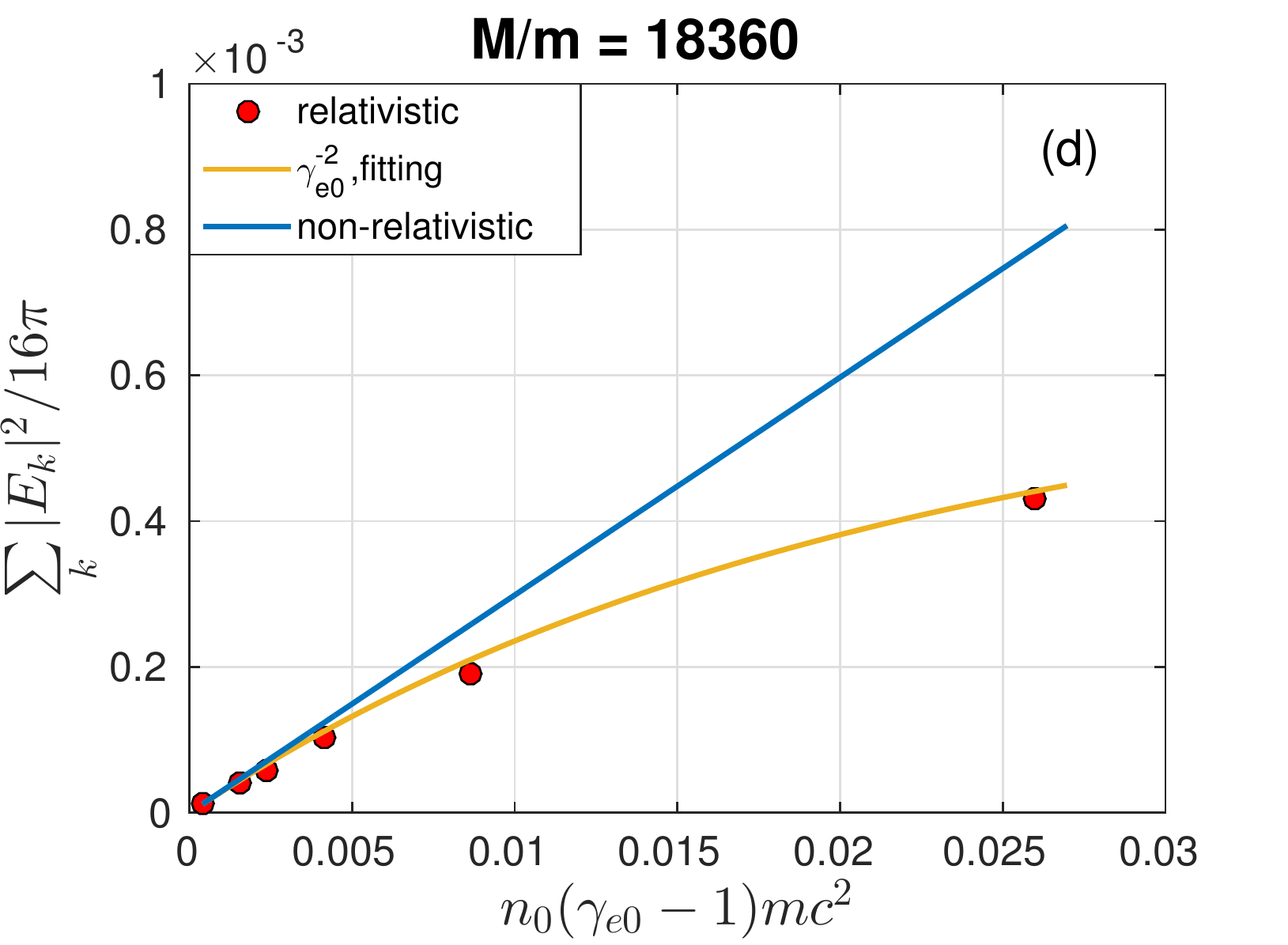}
	\label{fig:fig10(4)}
	}
	\caption{Evolution of electrostatic energy density with initial drift kinetic energy density for the mass ratio (\ref{fig:fig10(1)}) 500, (\ref{fig:fig10(2)}) 1000, (\ref{fig:fig10(3)}) 1836, (\ref{fig:fig10(4)}) 18360}.
	\label{fig:fig10}
	\end{figure}

\begin{figure}
\label{fig:fig8}
\includegraphics[scale=0.8]{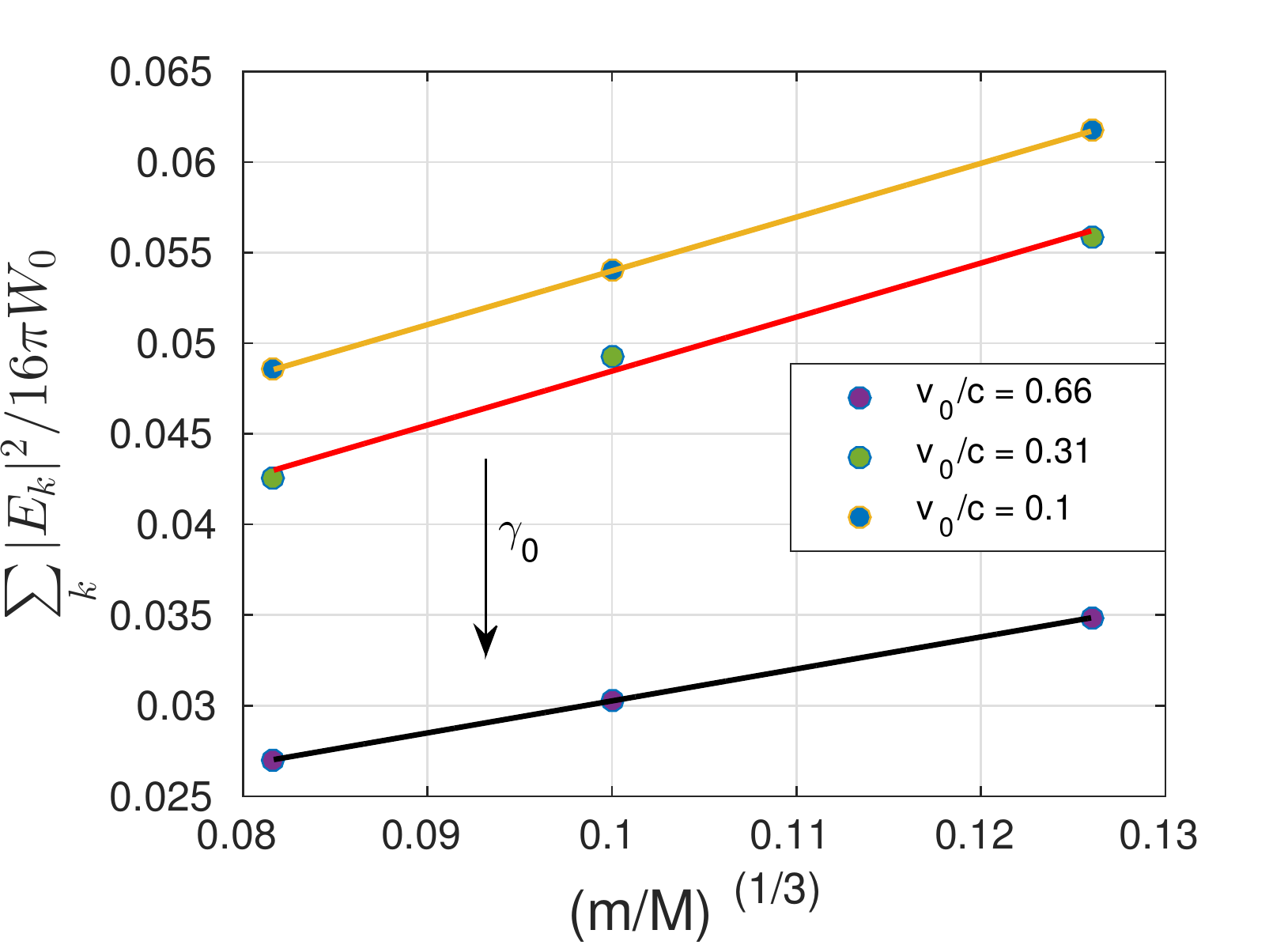}
\caption{Figure shows variation of $\sum\limits_{k}|E_{k}|^{2}/16 \pi W_{0}$ with mass ratio for different initial drift velocities.}
\end{figure}

Following an argument similar as above, in the relativistic case the growth rate ($\gamma/{\omega}_{pe}$) maximizes at the resonant wave number given as $kv_{0} \sim \omega_{pe}/\gamma^{3/2}_{e0}$, which also sharply drops for small changes in the drift velocity; the width of the $\gamma/\omega_{pe}$ vs $kv_{0}/\omega_{pe}$ curve may be estimated by replacing electron mass m by $m_{eff} = m \gamma^{3}_{e0}$ and $\omega_{pe}$ by $\omega^{'}_{pe} = \omega_{pe}/\gamma^{3/2}_{e0}$ in the weakly relativistic dispersion relation (equation \eqref{eq:11}) which leads to $\Delta (kv_{0}/\omega_{pe}) \sim \frac{1}{\gamma^{1/2}_{e0}} \left(\frac{m}{M}\right)^{1/3}$. Further the change in electron drift velocity at the saturation point may be estimated from the resonance condition as $\frac{k\Delta v_{0}}{\omega_{pe}} \sim -\frac{3}{2} \frac{\omega}{\gamma^{5/2}_{e0}} \Delta\gamma_{e0}$ implying that $\frac{k\Delta v_{0}}{\omega_{pe}}$ scales with relativistic factor $\gamma_{e0}$ as $\frac{k\Delta v_{0}}{\omega_{pe}} \sim \frac{1}{\gamma^{5/2}_{e0}} \sim \frac{1}{\gamma^{2.5}_{e0}}$. We have verified this scaling in our simulations. Figure (\ref{fig:fig9}) shows the variation of $k \Delta v_{0}/\omega_{pe}$ with $\gamma_{e0}$ for mass ratio M/m = 1836. The dots represent the points obtained from simulation and the straight line fit shows a scaling as $k\Delta v_{0} \sim \frac{1}{\gamma^{2.8}_{e0}}$ which closely agrees with our back-of-the envelope estimate. Therefore the ratio $\frac{k \Delta v_{0}}{\Delta(kv_{0})}$ scales with $\gamma_{e0}$ as $\frac{k \Delta v_{0}}{\Delta(kv_{0})} \sim \gamma^{-2}_{e0}$. Now assuming Hirose's \cite{Hirose} [equation \eqref{eq:16}] to holds in the weakly relativistic limit, we note that the ratio of electrostatic field energy density at the saturation point to initial electron drift kinetic energy density scales with $\gamma_{e0}$ as 
\begin{equation} \label{eq:17}
\frac{|E|^{2}}{16 \pi W_{0}} \sim \frac{k\Delta v_{0}}{\Delta(kv_{0})} \left( \frac{m}{M} \right)^{1/3} \sim \frac{1}{\gamma^{2}_{e0}} \left( \frac{m}{M} \right)^{1/3}
\end{equation}
We have verified the above scaling in our simulations. Figure (\ref{fig:fig10}) shows the variation of electrostatic field energy density at the saturation point with initial beam kinetic energy density for different mass ratios. The Yellow curve shows $\sim \frac{1}{\gamma^{2}_{e0}}$ scaling and the blue straight line shows the scaling for the non-relativistic case (presented here for comparison \cite{rupendra}). Figure (\ref{fig:fig8}) shows the variation of the ratio of electrostatic field energy density at the saturation point to initial electron beam kinetic energy density with electron to ion mass ratio for different initial electron drift velocities. The linear variation with $(m/M)^{1/3}$ again confirms equation \eqref{eq:17}.

\section{SUMMARY} \label{sum}
In this paper, we have studied the evolution and saturation of the relativistic Buneman instability in 1-D using a in-house developed particle-in-cell simulation code. Our results clearly show that relativistic effects have a stabilizing influence on the instability. The growth rates of unstable modes as measured from simulation show a good match with that obtained from fluid model. Further at the saturation point the electrostatic field energy density scales with the initial electron drift kinetic energy density as $\sim \frac{1}{\gamma^{2}_{e0}}$, where $\gamma_{e0}$ is the Lorentz factor associated with the initial electron drift velocity. This scaling closely matches our back-of-the envelope estimate based on Hirose's \cite{Hirose} analysis. A detailed derivation of the above scaling is currently in progress and will be reported elsewhere.
\begin{acknowledgments}
We thank Kushal Shah for useful discussions.
\end{acknowledgments}
\bibliography{rel_bune_v3_mod1}
\clearpage


\end{document}